\documentclass{ws-p8-50x6-00}

\newcommand{\x}{{\bf x}}
\newcommand{\y}{{\bf y}}

\newcommand{\kk}{{\bf k}}

\newcommand{\B}{{\bf B}}
\newcommand{\A}{{\bf A}}
\newcommand{\bdel} {{\mbox{\boldmath $\nabla$}}}
\newcommand{\balpha} {{\mbox{\boldmath $\alpha$}}}
\newcommand{\bPi} {{\mbox{\boldmath $\Pi$}}}

\begin{document}

\title{Constructing Confinement}

\author{E.S. Swanson}

\address{Department of Physics and Astronomy, University of Pittsburgh\\
 Pittsburgh PA 15260}

\author{A.P. Szczepaniak}

\address{
  Department of Physics and Nuclear Theory Center,
  Indiana University\\
  Bloomington, IN   47405-4202}

\maketitle

\abstracts{
The interaction between static quarks is derived by applying many-body
techniques to QCD in Coulomb gauge. The result is shown to be 
exact in the IR and UV limits, and agrees remarkably well with
lattice computations.
}

\section{Introduction}
One of the key impediments to understanding QCD at low energy is a quantitative
description of confinement. Here, we demonstrate how confinement may be constructed
with a nonperturbative many-body analysis of QCD in Coulomb gauge\cite{Szczepaniak:2001rg}.

The starting point is the QCD Hamiltonian in Coulomb gauge, as derived by Schwinger\cite{Schwinger}
and Christ and Lee\cite{Christ:1980ku,Zwanziger:1998ez}:

\begin{eqnarray}
\bar H &=&  \int d\x \psi^\dagger\left( -i \balpha\cdot\bdel + \beta m\right) \psi + 
{1\over 2}\int d\x \left( {\cal J}^{-1/2}\bPi {\cal J} \cdot \bPi {\cal J}^{-1/2} + \B\cdot\B\right)
\nonumber \\
 &-&g \int d\x \psi^\dagger \balpha\cdot \A \psi + H_c.
\end{eqnarray}

\noindent
where $H_C$ is the nonabelian version of the instantaneous Coulomb interaction:

\begin{equation}
H_C = {1\over 2}\int d\x d\y {\cal J}^{-1/2} \rho^a(x)
 {\cal J}^{1/2} K_{ab}(\x,\y;\A) {\cal J}^{1/2} \rho^b(y) {\cal J}^{-1/2}.
\label{hc2}
\end{equation}

\noindent
In this expression the total charge density is denoted by $\rho$, ${\cal J}$ is the Faddeev-Popov
determinant, ${\cal J} = {\rm det}(\nabla\cdot D)$, $D$ is the covariant derivative in adjoint
representation, $D^{ab} = \delta^{ab}\bdel - g  f^{abc} \A^c$, and $K$ is the nonabelian kernel

\begin{equation}
K_{ab}({\bf x},{\bf y};\A) \equiv \langle{\bf x},a|
 { g \over { \bdel\cdot {\bf D}}}(-\bdel^2)
 { g \over { \bdel\cdot{\bf D}}}|{\bf y},b\rangle,
\label{ck}
\end{equation}

\noindent
We shall make use of the inverse Faddeev-Popov operator, $d = g/(\bdel\cdot D)$ in the
following.

Coulomb gauge is particularly efficacious in describing QCD in the confinement region
because all degrees of freedom are physical (no gauge constraints need be applied) and the
instantaneous interaction provides an excellent starting point for the bound state problem.
Furthermore, it is easy to see that the Coulomb interaction {\it must yield confinement} 
for static quarks because transverse gluon exchange is suppressed in the heavy quark limit.
Thus it is clear that determining the behaviour of $K$ is of central importance to developing
an understanding of low energy QCD.

\section{IR and UV Behaviour of QCD in Coulomb gauge}

The evaluation of the Coulomb interaction is a nontrivial endeavour! However, the
problem simplifies dramatically if one is willing to focus on the leading infrared (IR)
behaviour of the theory (which, of course, is where confinement lies). To understand this
statement it is useful to recall an old story in many-body physics.

A naive attempt to compute the ground state energy of an electron gas
(with an assumed positive background) at higher order is confounded by IR divergences
arising from the instantaneous (abelian) Coulomb interaction. This problem was resolved
many years ago by Bohm and Pines\cite{BP} who showed that summing the leading IR-divergent
diagrams (the ring diagrams) effectively replaces the divergent Coulombic $1/q^2$ behaviour with a
convergent $1/q^2 \epsilon(\omega,k)$ behaviour.
In fact, the sum has a wholly different analytic structure than any finite order in
perturbation theory can produce.
We now demonstrate that a similar situation occurs in the nonabelian case. 

The vertices of QCD in Coulomb gauge are shown in Figure 1  (there are additional vertices
where $n$ transverse gluons are emitted from the Coulomb interaction). We denote these vertices
by $\alpha$ and call the number of vertices of a given type in a time-ordered diagram, $n_\alpha$.
We anticipate that the Coulomb kernel $K$ does indeed cause an IR divergence when inserted in a 
typical diagram.
We denote the divergence due to $K$ by $\lambda$ (for concreteness, one may
consider $\lambda$ as an IR regulator). A general diagram
then has an IR divergence of order $\lambda^M$ where $M = D_K - D_g - D_q$. Here $D_K$, $D_g$, 
$D_q$, and $D_d$  represent the number of Coulomb, gluon, quark, and Faddeev-Popov propagators 
respectively.  The following relationships may be easily  verified\cite{Swift:za}:
$2D_K = n_{ggK} + n_{qgK} + n_{Kgd}$, $2D_d = n_{Kgd} + 2 n_{dgd} + n_{qqd}$,
and $n_{Kgd} = n_{ggd} + n_{qqd}$; employing these leads to 
$M = 1 + 2D_K +D_d - \sum_\alpha n_\alpha$, and finally

\begin{equation}
M = 1 - n_{4g} - n_{3g} - n_{qqg}
\end{equation}

This equation is of central importance -- no diagrams are more divergent than $K$ itself, and
increasing the number of four gluon, three gluon, or quark-gluon vertices decreases the
degree of divergence. Thus {\it the leading IR behaviour of QCD is contained in matrix elements 
of the Hamiltonian itself}  -- specifically it is not in iterations of $H$.

\begin{center}
\begin{figure}[h]
\epsfxsize=18pc 
\epsfbox{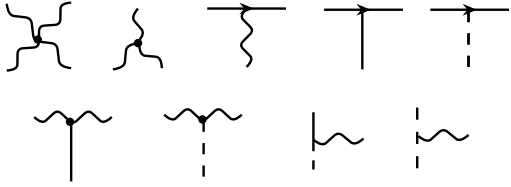} 
\caption{Vertices in Coulomb gauge QCD. The solid line is a $K$ propagator
while the dashed line represents a $d$ propagator.}
\end{figure}
\end{center}

Before applying these results we consider the ultraviolet (UV) behaviour of QCD
in Coulomb gauge. It is well known\cite{Feinberg:1977rc} that the beta function of
QCD may be obtained directly from the Coulomb interaction (there is no need to consider
loop corrections to vertices). Thus an accurate evaluation of $K$ is sufficient to determine
both the long distance and short distance behaviour of the static quark confinement potential!

\section{Coupled Gap Equations}

The previous observations simplify the evaluation of the Coulomb kernel: the leading IR divergent
diagrams coincide with those of the rainbow ladder approximation. Furthermore, vertex corrections
to the rainbow ladder Dyson equation are suppressed in the IR limit (see Ref. [1] for more details).
Thus $K$ may be evaluated by solving a relatively simple nonlinear integral equation. 

An important additional step needs to be taken before evaluating $K$. Diagrams are all evaluated
with respect to an implicit vacuum, normally taken to be trivial. However, a  trivial vacuum is not
likely to be a good starting point for describing confinement. Thus we choose to introduce extra
freedom into the problem with the aid of a variational vacuum ansatz:

\begin{equation}
\Psi_0[\A] = \langle \A | \omega \rangle =  \exp\left[-{1\over 2} \int d\kk
   \A^a(\kk)\,\omega(k)\A^a(-\kk) \right].
\label{varvac}
\end{equation}

\noindent
Here, $\omega$ is an unknown function to be determined by minimizing the vacuum energy 
density.

The particular form of the trial vacuum is unimportant -- it merely serves as a method
to define the basis with which we choose to describe low energy QCD.  The
Fock space which is built on our variational vacuum consists of quasiparticles
-- constituent quarks and gluons. These degrees of freedom obey dispersion
relations with infrared divergences due to the long-range instantaneous Coulomb
interaction of the bare partons with the
mean field vacuum. This interaction makes colored objects infinitely
heavy thereby removing them from the physical spectrum.
However, color neutral states remain physical because the
infrared singularities responsible for the large self-energies
are canceled by infrared divergences responsible for the long-range
forces between the constituents. This is the origin of the ``colour confinement" 
phenomenon.

Having specified the vacuum (up to the function $\omega$) we are ready to evaluate
the Coulomb interaction. Since the result depends on $\omega$ we must also fix $\omega$
by minimizing the vacuum energy density (this is just the gap equation). However, the
vacuum energy depends on $\langle \Psi_0|H_C|\Psi_0\rangle$, which depends on $K$! Thus 
we must solve {\it coupled} equations for the gap and for $K$. Doing so yields a result
for $\omega$ which may then be used to evaluate $K$ -- the equations themselves determine
the gluonic (and hence quark) quasiparticle interaction.

\begin{figure}[h]
\epsfxsize=14pc 
\epsfbox{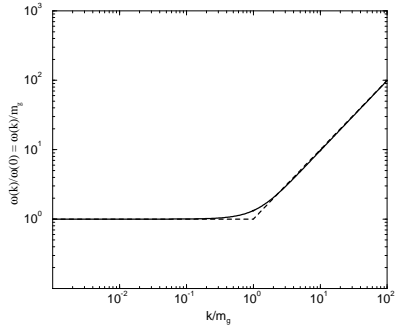} 
\caption{ Comparison of an analytical approximation to $\omega(k)$
  (dashed line) and the full numerical solution (solid line).}
\end{figure}

The details of this procedure are described in Ref. [1]. The result for $\omega$ is shown
in Figure 2. The data are presented with respect to a scale $m_g = \omega(0)$ which must be determined
by comparison with experiment (recall that this is full QCD, the theory needs to be
regulated and renormalized; the existence of $m_g$ is due to dimensional transmutation).
Finally, the Coulomb interaction may be evaluated. The result (after numerical Fourier 
transforming) is shown in Figure 3. Taking lattice data as ``experiment" allows us to
determine $m_g$ in terms of the lattice scale (the Sommer parameter). One finds $m_g \approx 600$
MeV.

\begin{figure}[h]
\epsfxsize=16pc 
\epsfbox{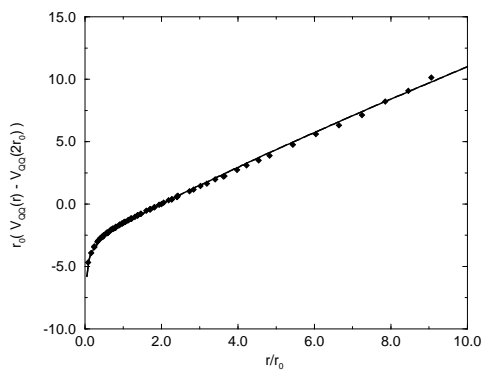} 
\caption{ Static $Q{\bar Q}$ ground state potential.
 The solid line is the numerical solution, 
 data are taken from Ref.~[8].}
\end{figure}

Finally, an examination of the  UV limit of the coupled gap equations reveals that
12/11 of the one loop  beta function is recovered. The reason is that the vacuum expectation
value of the Coulomb interaction does not include the gluon loop contribution to the 
running coupling. This is generated by iterating the Hamiltonian and may be incorporated
if desired. Figure 3 makes it clear that this will have little phenomenological effect.

\section{Conclusions}

Coulomb gauge allows dramatic simplification of the confinement (and bound state) problem
of QCD. Confinement must be contained in the nonabelian Coulomb interaction. We have shown
that the combination of allowing modest flexibility in the vacuum and a judicious evaluation
of the Coulomb kernel is sufficient to derive the lattice Wilson potential to great 
accuracy. 

Unfortunately, it is difficult to make cartoons which explain `why' confinement 
occurs in this formalism. Nevertheless it is tempting to speculate that the origin of confinement is 
related to the Gribov problem -- the nonuniqueness of Coulomb gauge. A simple practical
resolution of the Gribov problem is to restrict the set of gauge configurations to those
which form an absolute minimum of a functional in gauge transformation space (it is
$F[u] = {\rm Tr}\int d^3x (\A^u)^2$ where  $\A^u = u\A u^\dagger - u\bdel u^\dagger$).
The resulting set is called the fundamental modular region. It may be shown\cite{Zwanziger:1998ez}
that this
region is convex, includes the origin, and intersects the Gribov horizon (defined as the location in gauge
space where the Faddeev-Popov operator vanishes). Two factors of the inverse Faddeev-Popov
operator appear in the Coulomb interaction, and hence gauge configurations near the 
horizon are naively strongly suppressed. However the horizon forms a boundary in an
infinite dimensional space and therefore carries a vast entropy. It is possible that these
effects cancel, leaving a strong IR enhancement in the inverse Faddeev-Popov operator, $d$, and
therefore  in $K$.  Details of this picture and phenomenological implications are being
examined.

\section*{Acknowledgments}
ES would like to thank the organizers of the Lake Louise Winter
Institute for providing a truly marvelous venue for discussing
physics and Dean Karlen and Francois Corriveau for organizing the first annual LLWI
hockey game.
This work was supported by DOE under contracts
DE-FG02-00ER41135,  DE-AC05-84ER40150 (ES), and DE-FG02-87ER40365 (AS).


\begin{thebibliography}{99}

\bibitem{Szczepaniak:2001rg}
A.~P.~Szczepaniak and E.~S.~Swanson,
Phys.\ Rev.\ D {\bf 65}, 025012 (2002).

\bibitem{Schwinger}
J.~Schwinger, Phys.\ Rev.\ {\bf 127}, 324 (1962).

\bibitem{Christ:1980ku}
N.~H.~Christ and T.~D.~Lee,
Phys.\ Rev.\ D {\bf 22}, 939 (1980).

\bibitem{Zwanziger:1998ez}
D.~Zwanziger,
Nucl.\ Phys.\ B {\bf 485}, 185 (1997).


\bibitem{BP} D. Bohm and D. Pines, Phys. Rev. {\bf 42}, 609 (1953).

\bibitem{Swift:za}
A.~R.~Swift,
Phys.\ Rev.\ D {\bf 38}, 668 (1988).

\bibitem{Feinberg:1977rc}
F.~L.~Feinberg,
Phys.\ Rev.\ D {\bf 17}, 2659 (1978). See also T.D. Lee, {\sl Particle Physics and Introduction
to Field Theory}, (Harwood Academic, New York, 1981).

\bibitem{JKM}
K.J. Juge, J. Kuti, and C.J. Morningstar,  Nucl. Phys. Proc. Suppl. {\bf 63}, 326 (1998).


\end{thebibliography}
\end{document}